\documentclass[longbibliography,onecolumn,preprintnumbers,superscriptaddress,nofootinbib]{revtex4}
\usepackage{graphicx}
\usepackage{amssymb}
\usepackage{color}
\usepackage{enumerate}

\newcommand{\be}{\begin{eqnarray}}
\newcommand{\ee}{\end{eqnarray}}

\newcommand{\beaa}{\begin{equation} \begin{array}{ll}}
\newcommand{\eeaa}{\end{equation} 	\end{array} }

\usepackage{graphicx}
\usepackage{mathrsfs}
\usepackage{float}
\usepackage{multirow}
\usepackage{booktabs}

\usepackage{float}

\begin{document}

\title{Galactic rotation curves in Einstein's conformal gravity} 

\author{Qiang Li} 
\affiliation{Department of Physics, Harbin Institute of Technology, Harbin 150001, China}
\affiliation{Department of Physics, Southern University of Science and Technology, Shenzhen 518055, China} 
\author{Leonardo Modesto}
\email{lmodesto@sustc.edu.cn} 
\affiliation{Department of Physics, Southern University of Science and Technology, Shenzhen 518055, China} 


\begin{abstract}
We show quantitatively that an exact solution of Einstein's conformal gravity can explain very well the galactic rotation curves for a sample of $104$ galaxies without need for dark matter or other exotic modification of gravity. The metric is an overall rescaling of the Schwarzschild-de Sitter spacetime as required by Weyl conformal invariance, which has to be spontaneously broken, and the velocity of the stars depends only on two universal parameters determined on the base of the observational data.

\end{abstract}

\maketitle

\section{Introduction}

One of the greatest mysteries in cosmology in our days is the ``dark matter's'' or ``dark gravity's" puzzle. Indeed, in order to take into account of all the observational evidences (galactic rotation curves, structure formation in the universe, CMB spectrum, bullet cluster) we need to somehow modify the right or the left side of the Einstein's field equations. In this paper we do not pretend to provide a definitive answer to the long standing question of what dark matter is, but we want to make known an extremely interesting result based on some previous work by Mannheim. Therefore, we here concentrate on only one of the above listed issues, 
namely the galactic rotation curves. 

In this paper we assume the ``dark gravity'' paradigma and we try to explain the anomalous behaviour of stars in galaxies in the contest of gravitational theories invariant under conformal transformations (see below). 
The analysis here reported is universal and apply to any conformally invariant theory that has the Schwarzschild-de Sitter metric as an exact solution. However, for the sake of simplicity we will focus on Einstein's conformally gravity in presence of cosmological constant. The theory is specified by the following general covariant action functional,
\be
 S=\int \! d^4x \sqrt{-\hat{g}} \, \left(\phi^2\hat{R}+6\hat{g}^{\mu\nu}\partial_\mu\phi \partial_\nu \phi- 2 f \phi^4 \right)
 \label{CEG} \, , 
\ee
which is defined on a pseudo-Riemannian spacetime Manifold $\mathcal{M}$ equipped with a metric tensor field $\hat{g}_{\mu\nu}$, a scalar field $\phi$ (the dilaton), and it is also invariant under the following Weyl conformal transformation:
\be 
\hat{g}^\prime_{\mu\nu}=\Omega^2\hat{g}_{\mu\nu} \, ,  \quad \phi^\prime=\Omega^{-1}\phi \, ,
\ee
where $\Omega(x)$ is a general local function. 
In (\ref{CEG}) $f$ is a dimensionless constant that has to be selected extremely small in order to have a cosmological constant compatible with the observed value. 
The Einstein-Hilbert gravity is recovered whether the Weyl conformal invariance is broken spontaneously in exact analogy with the Higgs mechanism in the standard model of particle physics (for more details we refer the reader to \cite{bambi2017spacetime,chakrabarty2018unattainable}). One possible vacuum of the theory (\ref{CEG}) (exact solution of the equations of motion) is 
$\phi = {\rm const.} = 1/\sqrt{16 \pi G}$, together with $R_{\mu\nu} \propto \hat{g}_{\mu\nu}$. 
Therefore, replacing $\phi = 1/\sqrt{16 \pi G} + \varphi$ in the action and using the conformal invariance to eliminate the gauge dependent degree of freedom $\varphi$
, we 
finally get the Einstein-Hilbert action in presence of the cosmological constant,
\be
S_{\rm EH} = \frac{1}{16 \pi G} \int d^4 x \sqrt{-\hat{g}} \, \left(\hat{R} - 2 \Lambda \right) ,
\label{EH}
\ee
where $\Lambda$ is consistent with the observed value for a proper choice of the dimensionless parameter $f$ in the action (\ref{CEG}). 
 Ergo, Einstein's gravity is simply the theory (\ref{CEG}) in the spontaneously broken phase of Weyl conformal invariance \cite{bambi2017spacetime}. 
 
Besides the constant vacuum, if a metric $\hat{g}_{\mu\nu}$ is an exact solution of the equations of motion thus it is also the rescaled spacetime with a non trivial profile for the dilaton field, namely 
 \be
 \hat{g}_{\mu\nu}^*=S(r) \hat{g}_{\mu\nu} \, \quad \phi^* = S(r)^{-1/2} \, \phi . 
 \label{rescaling}
 \ee
Therefore, in this paper we push further the relevance of conformal symmetry in the real world on the footprint of previous works that were mainly concerning the singularity issue \cite{bambi2017spacetime,chakrabarty2018unattainable}. Contrary to the latter papers, we here focus on a non asymptotically flat rescaling on the Schwarzschild metric (or Schwarzschild-de Sitter metric) in order to explain the data for the galactic rotation curves with a minimal and universal choice of only two free parameters that the reader will meet later on in the paper. Finally, in order to give physical meaning to the metric (\ref{rescaling}) the conformal symmetry has to broken spontaneously to a particular vacuum specified by the function $S(r)$.


\section{The galactic geometry}
As explained in the introduction, given an exact solution of Einstein's conformal gravity, any rescaled metric is an exact solution too, if the metric is accompanied by a non-trivial profile for the dilaton. Therefore, we here consider the following conformal rescaling of the Schwarzschild-de Sitter spacetime, 
\be
&& ds^2= Q^2(x) \left[ -  \left(1-\frac{2GM_{\bigodot}}{c^2 x} - \frac{\Lambda}{3} x^2 \right) c^2 dt^2+\frac{dx^2}{1-\frac{2GM_{\bigodot}}{c^2 x} - \frac{\Lambda}{3} x^2}+x^2(d\theta^2+\sin^2\theta d\phi^2) \right] \, , \\
&& Q(x) = \frac{1}{1- \frac{\gamma*}{2} x } \, ,
\label{Qmetric}
\ee
where we identified $x$ with the radial coordinate. The reason of the particular rescaling $Q(r)$ will be clarify shortly. 
Now we perform a coordinate transformation in order to express the metric in the usual radial Schwarzschild coordinate, which identifies the physical radius of the two-sphere. Moreover, the transformation of coordinates should also be compatible with the usual relation $g_{00} = -1/g_{11}$. These requirements uniquely fix the rescaling factor. Indeed, in the new radial coordinate $r$, defined by 
\be
x=\frac{r}{1 + \gamma^* \frac{r}{2} } \, , 
\ee
the metric reads: 
 \be
&& ds^2= -  Q^2(r) \left( 1-\frac{2GM_{\bigodot} Q(r)}{c^2 r} - \frac{\Lambda}{3} \frac{r^2}{Q^2(r)} \right) c^2 dt^2+\frac{dr^2}{Q^2(r) \left( 1-\frac{2GM_{\bigodot} Q(r)}{c^2 r} - \frac{\Lambda}{3} \frac{r^2}{Q^2(r)} \right)} + r^2 (d\theta^2+\sin^2\theta d\phi^2)  \, , 
\nonumber \\
&& Q(r) = 1+ \frac{\gamma*}{2} r  \qquad (\mbox{notice that} \,\, x = r/Q) \, .
\label{Qmetricr}
\ee
It deserve to be notice that $x = r/Q$ and, therefore, the asymptotic contribution to the metric due to the cosmological constant is independent on the rescaling $Q(r)$. 
Moreover, it is interesting that any rescaling which differs from the one in (\ref{Qmetric}) is not consistent with the two requirements above. Therefore, in the infinite class of exact solutions conformally equivalent to the Schwarzschild-de Sitter metric, there is only one geometry non-asymptotically flat consistent with $g_{00} = -1/g_{11}$ and two-dimensional transverse area $4 \pi r^2$.

\subsection*{The local contribution to the stars' velocity}

Inside a galaxy the metric (\ref{Qmetricr}) represents the geometry around any star, and the coordinate $r$ has the usual interpretation of radial coordinate. Therefore, we can easily compute the Newtonian gravitational potential 
in which any other star moves in, i.e.
\be
V &=& -\frac{1}{2}(g_{00}(r)+c^2)
=  -\frac{1}{2} c^2 \left[ -Q^2(r) \left( 1-\frac{2GM_{\bigodot}Q(r)}{c^2 r} \right)+1 \right] \nonumber\\
&=& \frac{\gamma^* c^2}{2} r
+\frac{(\gamma^* )^2c^2}{8} r^2 
-\frac{GM_{\bigodot}}{r} \left( 1+\frac{3\gamma^*}{2} r 
+\frac{3(\gamma^*)^2}{4} r^2 + \frac{(\gamma^*)^3}{8} r^3\right) \, .
\label{potential}
\ee
At the galactic scale $\gamma^* r \ll 1$ \cite{mannheim2012fitting}\footnote{Notice that we can identify the radial coordinate $r$ with the radial physical length, namely 
$d \ell_r = \frac{dr}{1+ \epsilon} \approx dr$,
because $\epsilon$ is proportional to $G M/r$ or to integer powers of the product $\gamma^* \, r$ that are much smaller then $1$.}, hence a star of solar mass generates a Newtonian gravitational potential with a linear asymptotic correction in the galactic halo given by  
\begin{equation}\label{8}
	V = -\frac{GM_{\bigodot}}{r} + \frac{\gamma^* c^2}{2} r \, .
	\label{Gp}
\end{equation}
We have removed the contribution of the cosmological constant that we will reintroduce later. 

The next step is to take into account the contribution to the gravitational potential on a probe solar mass star in the halo due to all the other stars in the galaxy sample. In other words (or in Newtonian terms) we have to consider the resulting force acting on the probe star and due to its interaction with all the other stars in the same galaxy. 

For this purpose we have to use the distribution of the luminous matter in an thin disk galaxy whose surface brightness is consistent with the following profile,
\be
\Sigma(R) = \Sigma_0 \, e^{-R/R_0} \, , 
\ee 
where $R$ is the distance from galactic center in cylindrical coordinates, which should not be confused with the radial coordinate $r$ between two stars in spherical coordinates, $R_0$ is the disk scale length of the galaxy, and $\Sigma_0$ is defined  by means of the total luminosity of the galaxy $L = 2 \pi \Sigma_{0}^2$.

Therefore, we have to integrate equation (\ref{potential}) over the whole galaxy to finally get the rotation velocity for a star orbiting in the plane of the galactic disk. 
The result is the local galactic contribution to the velocity of a single star \cite{mannheim2006alternatives}, namely 
\begin{equation}\label{LOCAL}
v^2_{\rm LOC}(R)=\frac{N^*GM_{\bigodot}R^2}{2R_0^3}[I_0(x)K_0(x)-I_1(x)K_1(x)]+\frac{N^*\gamma^*c^2R^2}{2R_0}I_1(x)K_1(x) \, , 
\end{equation}
where here $x \equiv R/2R_0$ (this definition should not be confuse with the radial coordinate in (\ref{Qmetric})) and $N^*$ is the total number of solar mass stars in the galaxy, while $I_0$, $I_1$ are the modified Bessel functions of first kind and $K_0$, $K_1$ are the modified Bessel functions of second kind.

%

\begin{figure}
	\centering \includegraphics[height=8cm,width=8cm]{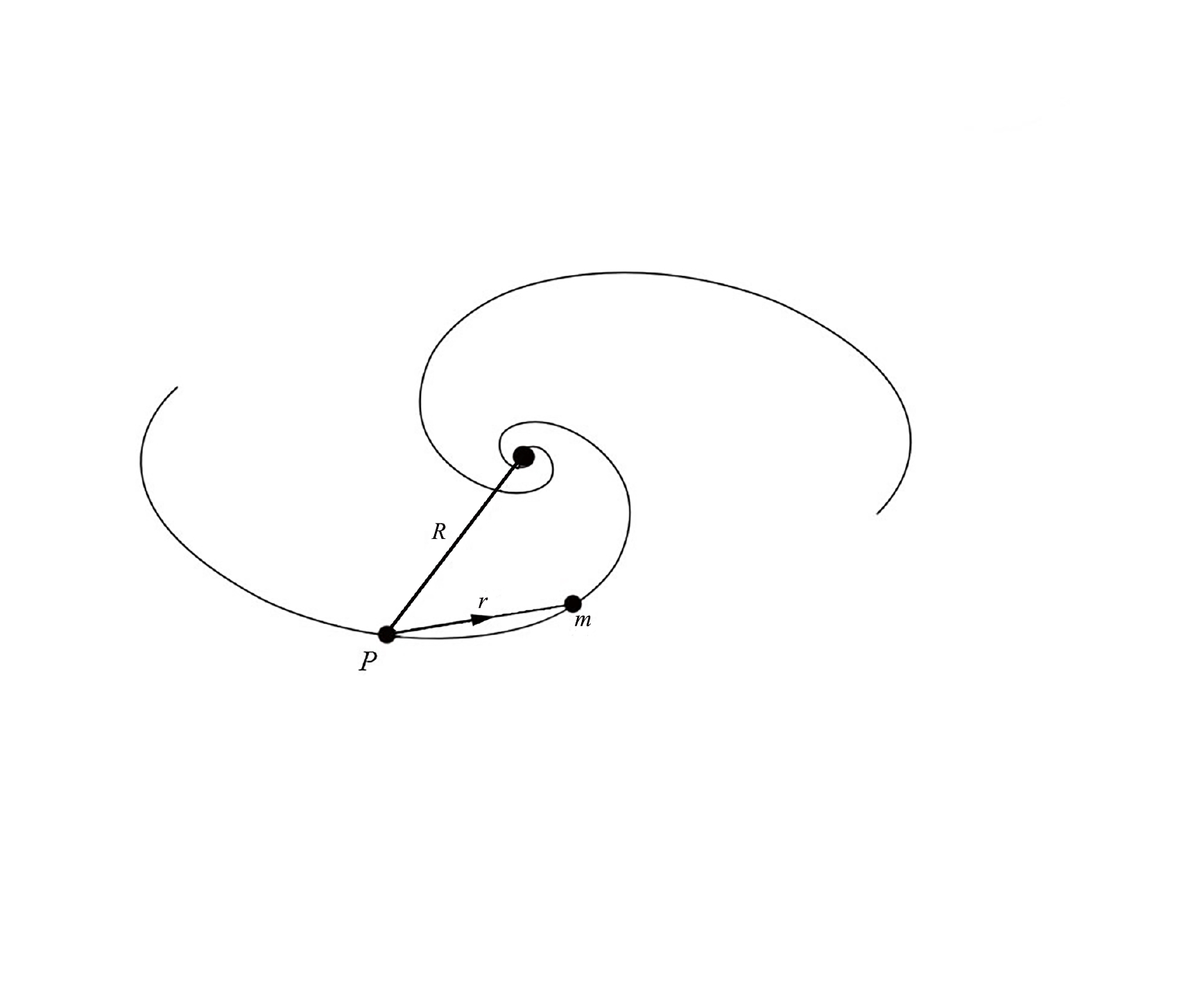} \hspace{0cm}
	\centering \includegraphics[height=8cm,width=8cm]{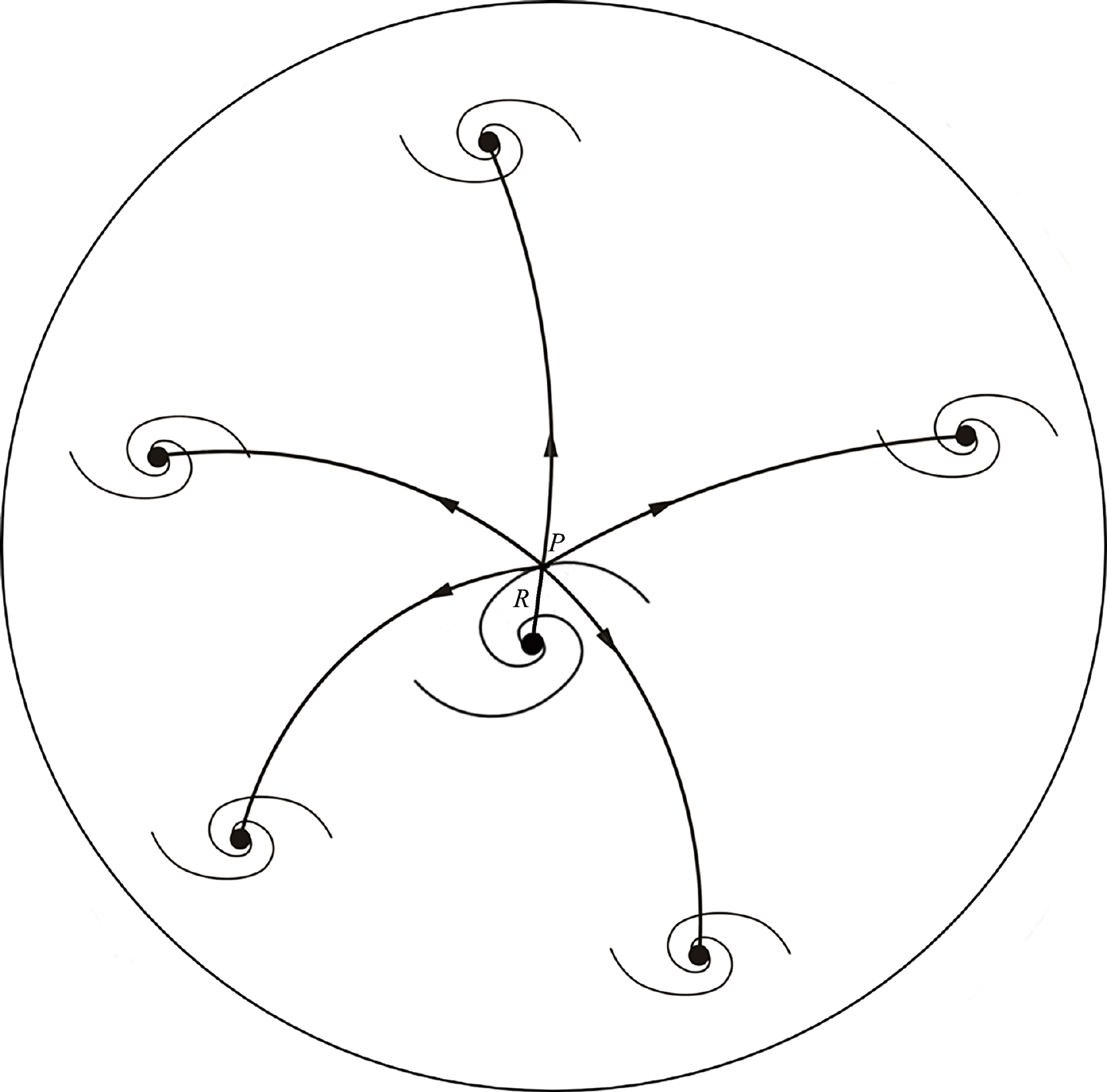}
	\caption{The contribution to the gravitational force acting on a star, in a generic point $P$ of a sample galaxy, consists of a local contribution due to the other stars in the same galaxy (picture on the left) and a global contribution due to the other galaxies in all the Universe (picture on the right).}
	\label{fig:1}
\end{figure}

%
%
%

\section{The Universe Geometry}



In the previous section we derived the local geometry surrounding any star in the galaxy and integrated over all the stars in the same galaxy to finally get the force acting on a probe star. In this section we would like to investigate whether the other galaxies of the whole universe can significantly have influence on the motion of the stars in a sample galaxy.

As painted in Fig.\ref{fig:1}, the probe star, labeled as a point $P$ in a sample galaxy, feels the gravitational attraction of all the other stars in the same galaxy, but also the effect of all the galaxies in the Universe due to the non homogeneous nature of the latter. Indeed, the latter global effect would vanish in a perfectly homogeneous universe because, saying in Newtonian terms,  the contributions to the total gravitational force coming from opposite directions erase each other.
Furthermore, we also have to take into account the global contribution due to the cosmological constant that we will consider later.
For the moment, the question we should find an answer is: what is the geometry due to all the other galaxies in the universe? First of all we observe that the metric (\ref{Qmetricr}) is not asymptotically flat and the linear contribution grows with the radial coordinate $r$, which makes a long distance interaction possible. Therefore, we can assume that the global contribution of the other galaxies in the Universe is well mimic by a metric very similar to (\ref{Qmetricr}). 
The last statement can be proved making a coordinate transformation that turns the metric (\ref{Qmetricr}) in a metric conformal to the Robertson-Walker (RW) spacetime \cite{mannheim2011impact}. Indeed, 
%
%
%
%
under the coordinate transformation $(t,r,\theta, \phi) \,\, \rightarrow \,\, (\tau, \rho,\theta, \phi)$, i.e. 
\be
	r = \frac{\rho}{(1 - \gamma_0 \rho/4)^2 } \, , \qquad 
	t = \int  \frac{d \tau}{a( \tau )}  \, , 
\ee
it has been proved in \cite{mannheim2012fitting,mannheim1989exact} that the following metric (we here consider only the large distance leading contribution to (\ref{Qmetricr}), and, for the sake of simplicity we temporarily omit the cosmological constant that we will reintroduce later) 
\begin{equation}
ds^2  = -(1+\gamma_0 r) c^2 dt^2 + \frac{dr^2}{1+\gamma_0 r} + r^2 d\Omega^{(2)}
\label{GG}
\end{equation}
turns into 
\be
 ds^2 & = & \frac{1}{a^2(\tau)}(\frac{1+\gamma_0\rho/4}{1-\gamma_0\rho/4})^2 \left[ -c^2d\tau^2+\frac{a^2(\tau)}{ ( 1-\gamma_0^2\rho^2/16 )^2} \left( d\rho^2+\rho^2d\Omega^{(2)} \right) \right] \nonumber \\
& \equiv & e^{2 \omega(\rho, \tau)} \left[ -c^2d \tau^2+\frac{a^2(\tau)}{(1-\gamma_0^2\rho^2/16 )^2} \left( d\rho^2+\rho^2 d\Omega^{(2)} \right) \right] \, ,
\label{dBHRW}
\ee
which is clearly conformal to a slightly inhomogeneous RW metric whether $\gamma_0$ is a very small parameter of inverse length dimension\footnote{In our case $\gamma_0 \sim 10^{-30} {\rm cm}^{-1}$, hence for $\rho$ of the order of the observable Universe we get 
$\gamma_0^2 \rho_{H}^2/16 \sim (10^{-30} {\rm cm}^{-1} \times 10^{28} {\rm cm} )^2/10 \sim 10^{-5}$, which is negligible respect to $1$ in the metric (\ref{dBHRW}).}. Indeed, the metric (\ref{dBHRW}) represents an homogeneous spacetime for $\gamma_0=0$. 

In short, the RW metric (\ref{dBHRW}) turns into (\ref{GG}) (or (\ref{Qmetricr})) when expressed in terms of coordinates used by an observer at rest in the center of a chosen galaxy.

In force of the above global analysis we can now come back to the geometry (\ref{Qmetricr}). Since the RW metric (\ref{dBHRW}) is isotropic and quasi-homogeneous, every point in the geometry can serve as the origin of the coordinates. Hence, we can place the origin of coordinates in the center of a given comoving galaxy, 
%
and we can identify $r$ ($r \sim \rho$)  with $R$ introduced in the previous section. Indeed, the radial coordinate $r$ labels any point in the galaxy (respect to its center) and in particular the position of a probe star whose coordinate distance from the center of the galaxy was in the previous section indicated with $R$. 
%
%
Notice that for the global contribution we do not have to integrate over the all galaxies in the Universe, but we have to  parametrize such effect with a constant $\gamma_0$. 

Therefore, the geometry that provides a global contribution to the velocity is again (\ref{Qmetricr}) in which we have to replace the coordinate $r$ with $R$
, namely 
\be
ds^2_{\rm global} & = & -  \left( Q^2(R)  - \frac{\Lambda}{3} R^2\right) c^2 dt^2+\frac{dR^2}{ \left( Q^2(R) - \frac{\Lambda}{3} R^2\right)} + R^2 (d\theta^2+\sin^2\theta d\phi^2)  \, , 
\nonumber \\
& \approx  & -  \left( 1 + \gamma_0 R  - \frac{\Lambda}{3} R^2\right) c^2 dt^2+\frac{dR^2}{ \left( 1+ \gamma_0 R - \frac{\Lambda}{3} R^2\right)} + R^2 (d\theta^2+\sin^2\theta d\phi^2)  \, , 
\nonumber \\
Q(R) & = & 1+ \frac{\gamma_0}{2} R   \, .
\label{Qmetricr2}
\ee
where we assumed $\gamma_0^2 \ll \Lambda$ in order to remove the contribution $\gamma_0^2 R^2/4$ relative to $\Lambda R^2/3$.  This assumption will be confirmed by our best fit that provides the value $\gamma_0 \sim 10^{-30} {\rm cm}^{-1}$, while $\Lambda =1.1\times 10^{-56} {\rm cm}^{-2}$ (see next section). 
Using again $V = -(g_{00}(r)+c^2)/2$ for the metric (\ref{Qmetricr}) we obtain the global contribution to the velocity of the stars, namely 
\be
v_{\rm GLOBAL}^2= \frac{\gamma_0c^2}{2} R - \frac{\Lambda c^2}{3} R^2 \, .
\label{GLOB}
\ee

%


%
%
%

\section{Fitting of the galactic rotation curves}

\indent Now collecting the local contribution (\ref{LOCAL}) and global contribution (\ref{GLOB}) in the weak gravity limit \cite{mannheim1997galactic} we end up with the following total contribution to the velocity of a probe star in a sample galaxy, 
\begin{equation}\label{15}
	v_{\rm TOT}^2=v_{\rm LOC}^2+\frac{\gamma_0c^2}{2} R - \frac{\Lambda c^2}{3} R^2 \, ,
\end{equation}
The two universal parameters $\gamma^*$ and $\gamma_0$ will be determined in the this section on the base of the data in Table (\ref{Tab01}) and (\ref{Tab02}) for a sample of $104$ galaxies.

We acquired all the parameters for the luminosities ($L$), disk scale lengths ($R_0$), and HI gas mass ($M_{HI}$) from two sources: a previous work \cite{mannheim2012fitting} and from the SPARC database \cite{lelli2016sparc}. The parameters of $5$ galaxies (namely: ESO 079-G014, UGC 5716, UGC 12632, UGC A442, and UGC A444) have been extracted from SPARC. For the rotation velocities ($v$), $76$ out of $104$ galaxies can be found in SPARC database, the data for the other $28$ galaxies are here listed: ESO0840411\cite{mcgaugh2001high}, ESO 1200211\cite{mcgaugh2001high}, ESO 1870510\cite{mcgaugh2001high}, ESO 2060140\cite{mcgaugh2001high}, ESO 3020120\cite{mcgaugh2001high}, ESO 3050090\cite{mcgaugh2001high}, ESO 4250180\cite{mcgaugh2001high}, ESO 4880490\cite{mcgaugh2001high}, UGC 4115\cite{mcgaugh2001high}, UGC 11454\cite{mcgaugh2001high}, UGC 11583\cite{mcgaugh2001high}, UGC 11616\cite{mcgaugh2001high}, UGC 11648\cite{mcgaugh2001high}, UGC 11748\cite{mcgaugh2001high}, UGC 11819\cite{mcgaugh2001high}, F 730-V1\cite{mcgaugh2001high},
 NGC 925\cite{de2008high}, NGC 3031\cite{de2008high}, NGC 3621\cite{de2008high}, NGC 3627\cite{de2008high}, NGC 4826\cite{de2008high}, NGC 959\cite{de2008mass}, NGC 7137\cite{de2008mass}, UGC 1551\cite{de2008mass}, NGC 4395\cite{de2006high}, UGC 477\cite{de2006high}, M 33\cite{warner1973high}, NGC 1560\cite{broeils1992mass}. 

Our sample of galaxies cover high surface brightness (HSB) galaxies, which have large $N^*$ and $\Sigma_0$ ($\Sigma_0\propto N^*/R_0^2$), low surface brightness (LSB) galaxies with small $\Sigma_0$, and dwarf galaxies with small $N^*$. 

From (\ref{15}) it follows that the rotation velocity curve will fall off at large distance because of the presence of a negative quadratic term. Thus, in Fig.\ref{fig:2}-\ref{fig:8}, one can see that our formula (\ref{15}) 
is in good agreement with the actual data. 

In the fitting, we employ a disk model with exponential decay to describe HI gas distribution, and we choose the ratio of the gas scale length to the optical disk scale length to be four because the gas extend well beyond the optical disk. 
We have also multiplied the HI gas mass by $1.4$ to include the primordial Helium.
As a result, we found good fits with two universal parameters fixed to the values: $\gamma^*=3.32\times10^{-39} {\rm m}^{-1}$, $\gamma_0=3.3\times10^{-28}{\rm m}^{-1}$.  
 
Few further technical comments are needed. 
For the galaxy NGC 3109, we corrected the HI surface density to the value $1.67$ instead of $1.4$ as pointed in \cite{begeman1991extended}. 
Following the argument in \cite{mannheim2012fitting}
we removed the first ten velocity-points for the galaxy NGC 4826 reported in \cite{de2008high}.
The galaxy NGC $2976$ was reported in \cite{simon2003high} to have a double disk structure. However, considering that the rotation velocity data in \cite{de2008high} is within a distance of $2.5$ kpc and most of the points are located at a distance less than $1.5$ kpc, we used the approximation of a single disk with a scale length equal to $1.2$ kpc.
The plots in the Fig.s \ref{fig:2}-\ref{fig:8} show the fitting rotation velocity curves (in km ${\rm sec}^{-1}$) as functions of the radial distance (in kpc) from the galactic centers . Moreover, in all plots the observative data with errors (in km ${\rm sec}^{-1}$) are also displayed for the all sample of 104 galaxies

\section{Conclusions}
In this paper we have selected a spherically symmetric solution of Einstein's conformal gravity in presence of cosmological constant compatible with very minimal assumptions: (i) the metric is written in Schwarzschild coordinates in which the two dimensional transverse area  
equals $4 \pi r^2$, (ii) the timelike and spacelike components of the metric satisfy the relation $g_{00} = - g_{11}$. This two requirements specify uniquely the metric in the infinite class of metrics conformally equivalent. 

The resulting geometry captures the local and global structure of the universe up to two constant parameters that can be set based on observational data. 
The first of the two parameters ($\gamma^*$) is related to the only ambiguity present in the metric describing the local geometry, while the second parameter ($\gamma_0$) appears in a similar metric, but now describing the global geometry 
due to all the other galaxies in the Universe. 
Indeed, the metric proposed in this paper when expressed in comoving coordinates takes the form of a slightly non-homogeneous Robertson-Walker geometry. 

\begin{figure}[H]
	\centering
	\includegraphics[height=20cm,width=\linewidth]{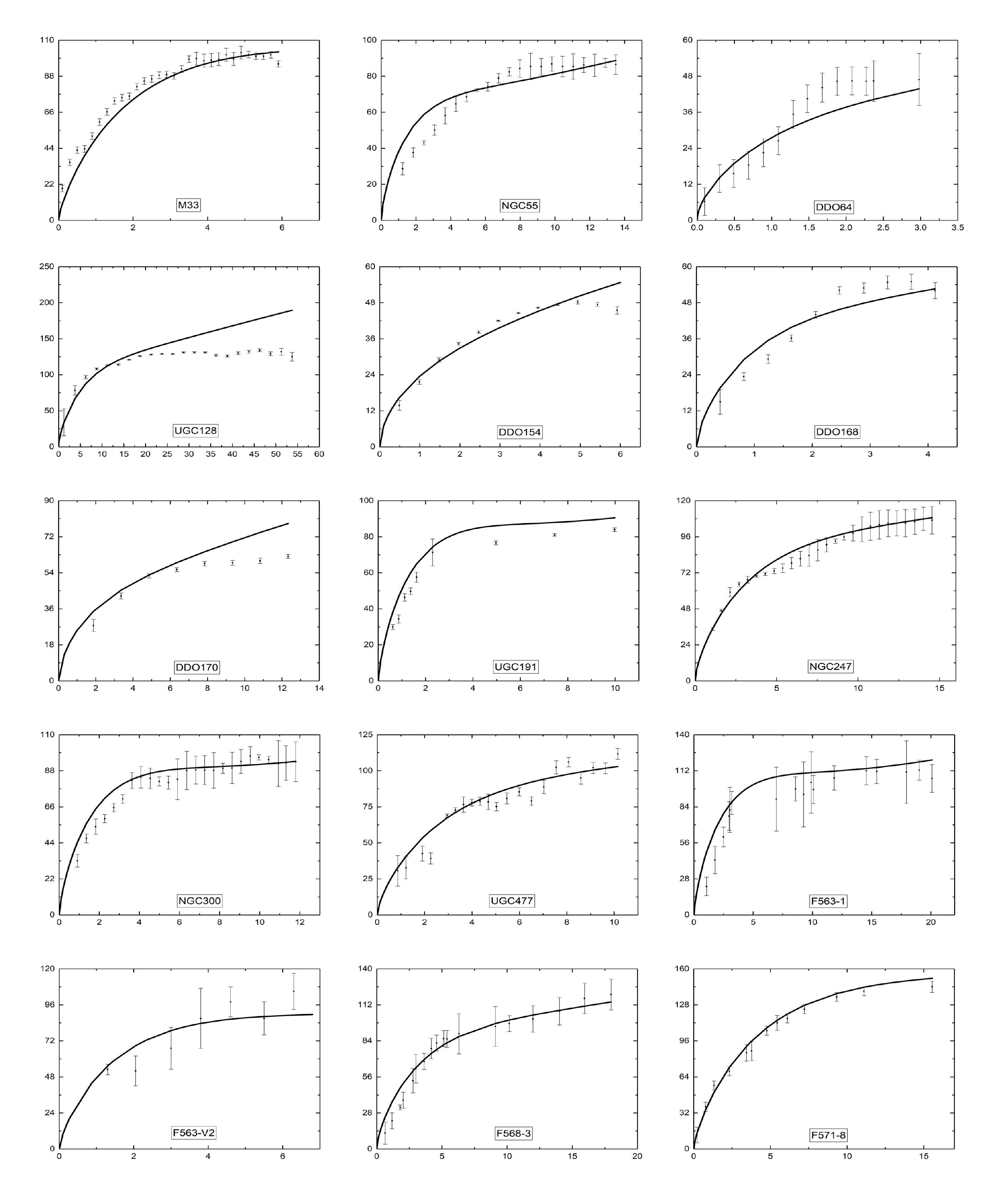}
	\caption{Fitting for the galaxies: M 33, NGC 55, DDO 64, UGC 128, DDO 154, DDO 168, DDO 170, DDO 191, NGC 247, NGC300, UGC 477, F563-1, F563-V2, F568-3, and F571-8.}
	\label{fig:2}
\end{figure}

\begin{figure}[H]
	\centering
	\includegraphics[height=20cm,width=\linewidth]{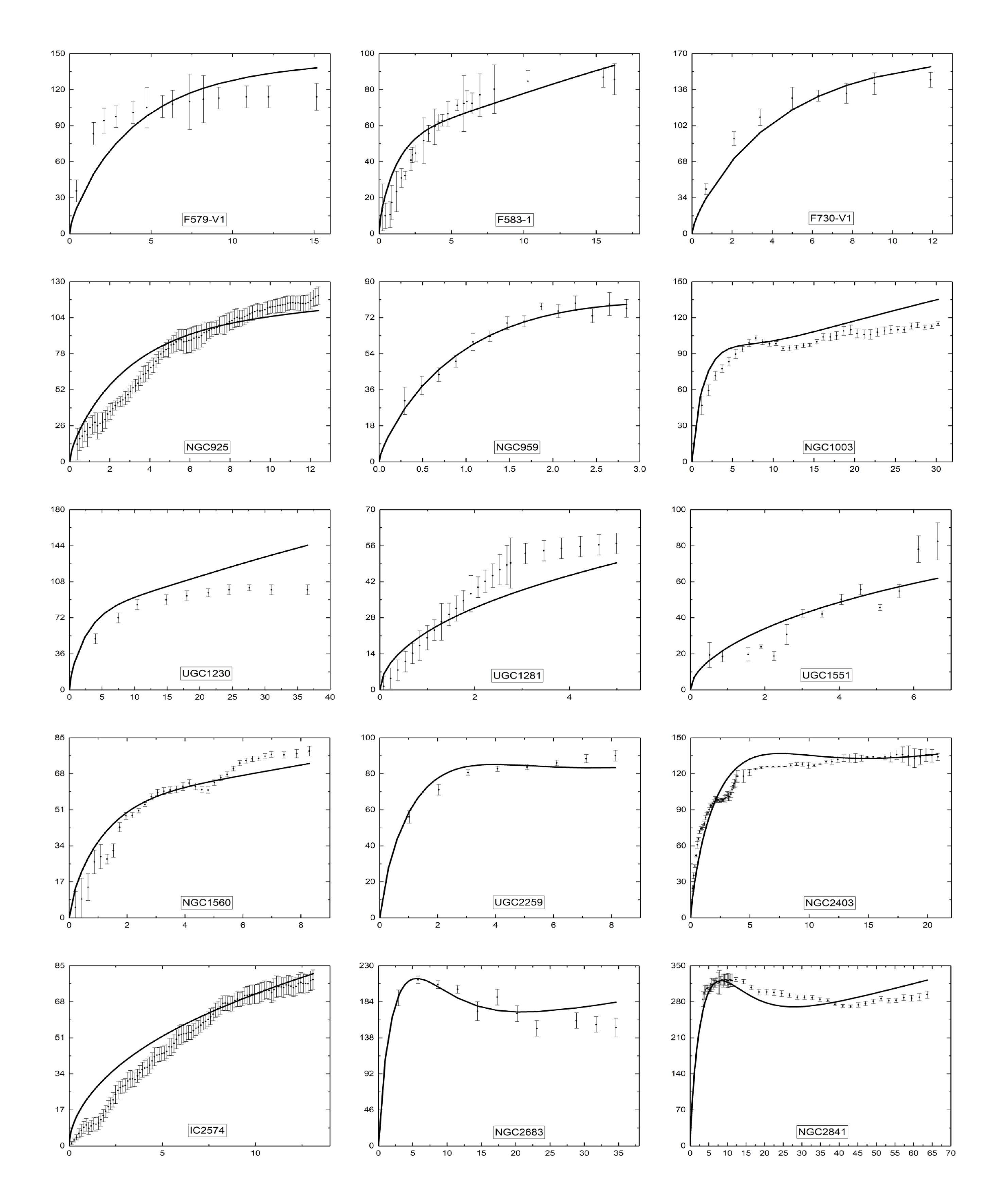}
	\caption{Fitting for the galaxies: F 579-V1, F 583-1, F 730-V1, NGC 925, NGC 959, NGC 1003, UGC 1230, UGC 1281, UGC 1551, NGC 1560, UGC 2259, NGC 2403, IC 2574, NGC 2683, and NGC 2841.}
	\label{fig:3}
\end{figure}

\begin{figure}[H]
	\includegraphics[height=20cm,width=\linewidth]{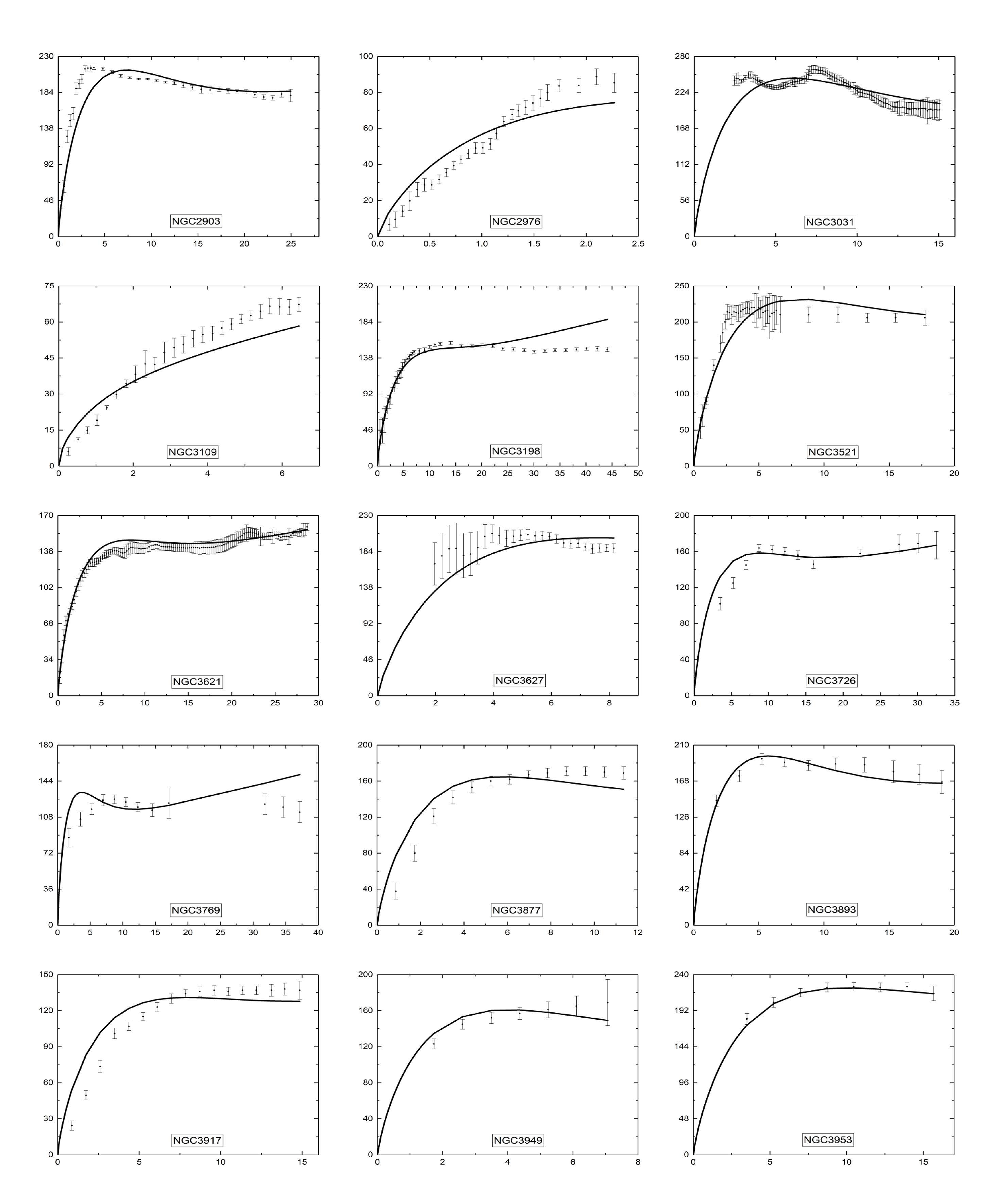}
	\caption{Fitting for the galaxies: NGC 2903, NGC 2976, NGC 3031, NGC 3109, NGC 3198, NGC 3521, NGC 3621, NGC 3627, NGC 3726, NGC 3769, NGC 3877, NGC 3893, NGC 3917, NGC 3949 and NGC 3953.}
	\label{fig:4}
\end{figure}

\begin{figure}[H]
	\centering
	\includegraphics[height=20cm,width=\linewidth]{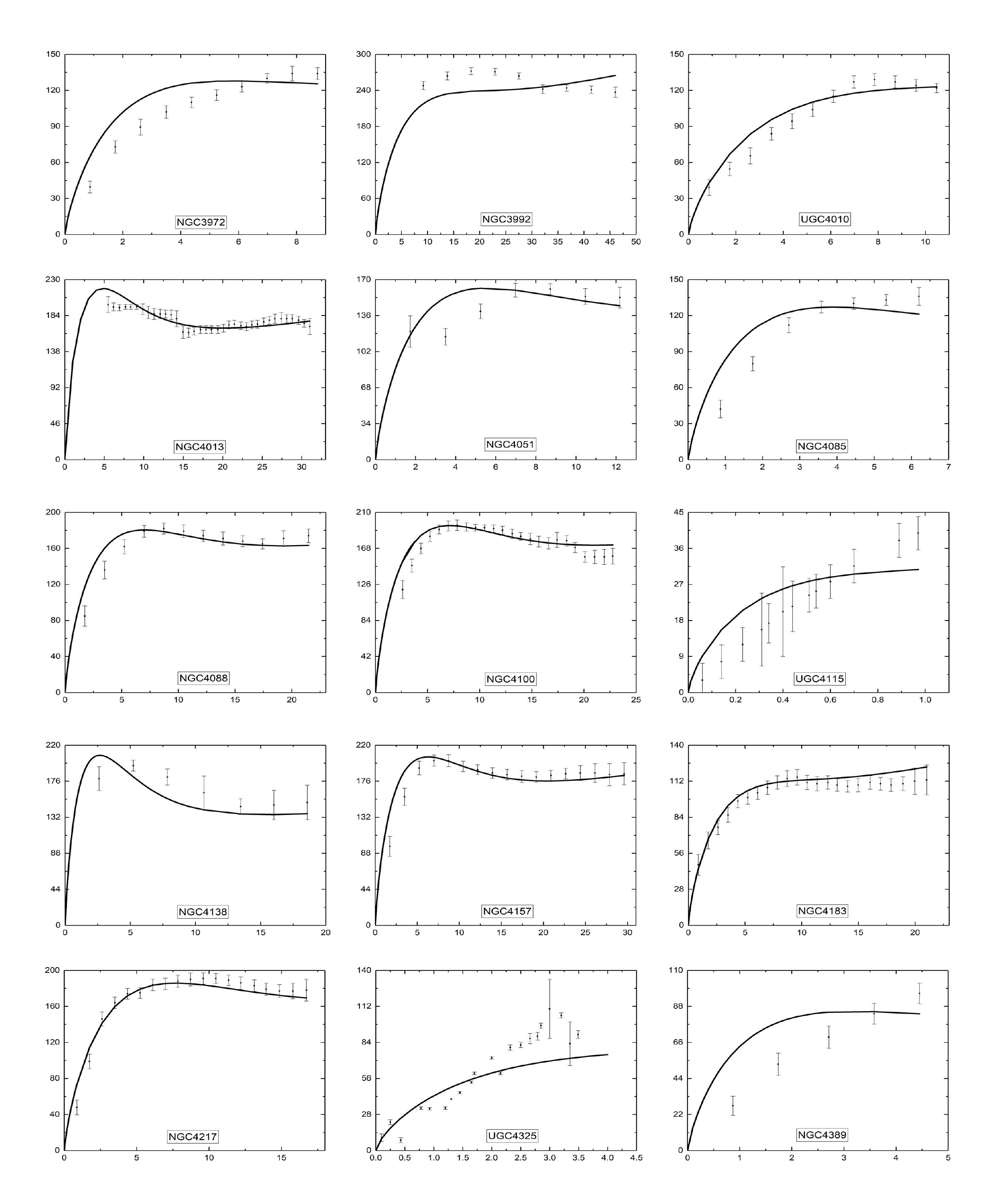}
	\caption{Fitting for the galaxies: NGC 3972, NGC 3992, UGC 4010, NGC 4013, NGC 4051, NGC 4085, NGC 4088, NGC 4100, UGC 4115, NGC 4138, NGC 4157, NGC 4183, NGC 4217, UGC 4325, and NGC 4389.}
	\label{fig:5}
\end{figure}

\begin{figure}[H]
	\centering
	\includegraphics[height=20cm,width=\linewidth]{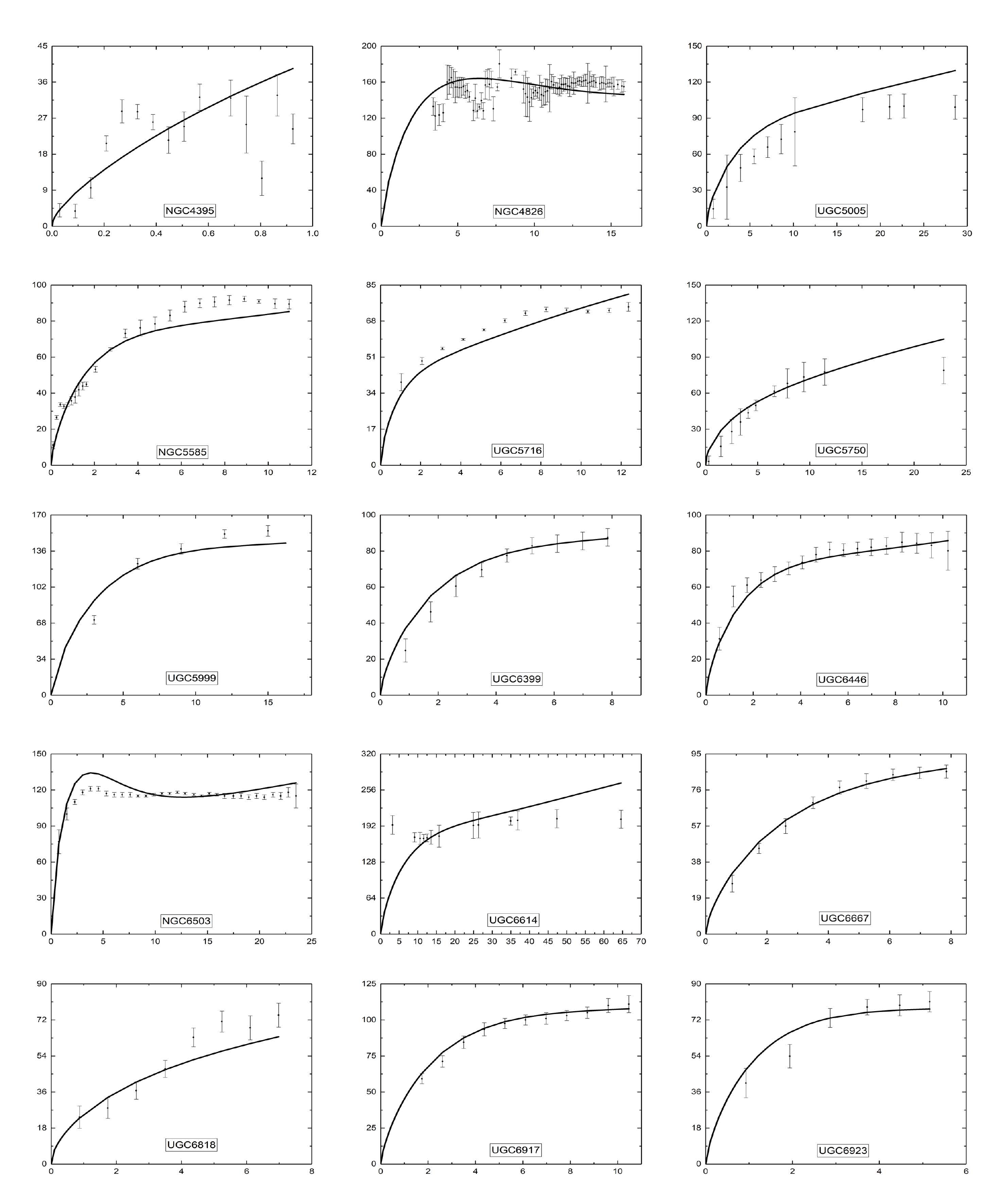}
	\caption{Fitting for the galaxies: NGC 4395, NGC 4826, UGC 5005, NGC 5585, UGC 5716, UGC 5750, UGC 5999, UGC 6399, UGC 6446, NGC 6503, UGC 6614, UGC 6667, UGC 6818, UGC 6917, and UGC 6923.}
	\label{fig:6}
\end{figure}

\begin{figure}[H]
	\centering
	\includegraphics[height=20cm,width=\linewidth]{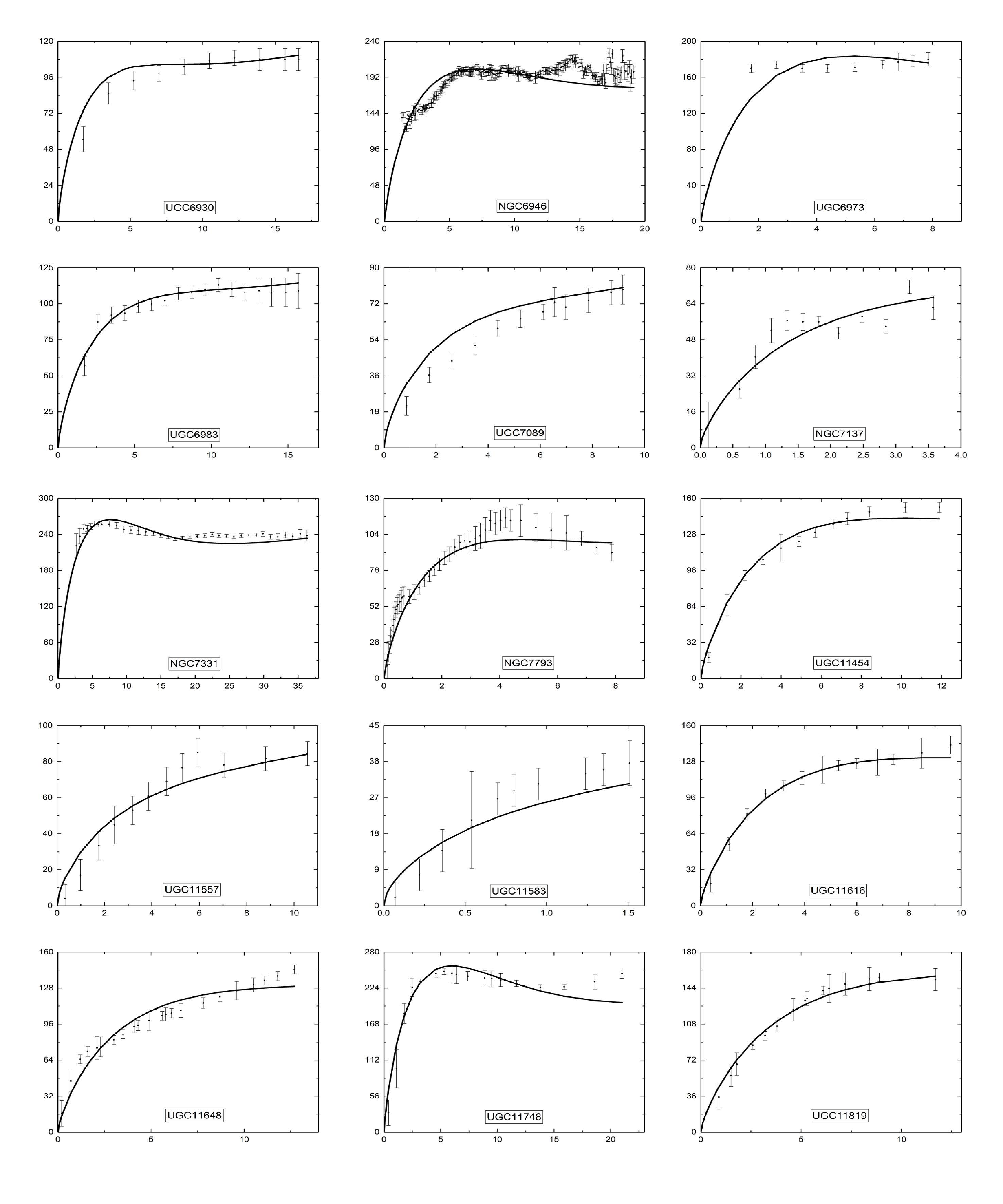}
	\caption{Fitting for the galaxies: UGC 6930, NGC 6946, UGC 6973, UGC 6983, UGC 7089, NGC 7137, NGC 7331, NGC 7793, UGC 11454, UGC 11557, UGC 11583, UGC 11616, UGC 11648, UGC 11748, and UGC 11819.}
	\label{fig:7}
\end{figure}

\begin{figure}[H]
	\centering
	\includegraphics[height=20cm,width=\linewidth]{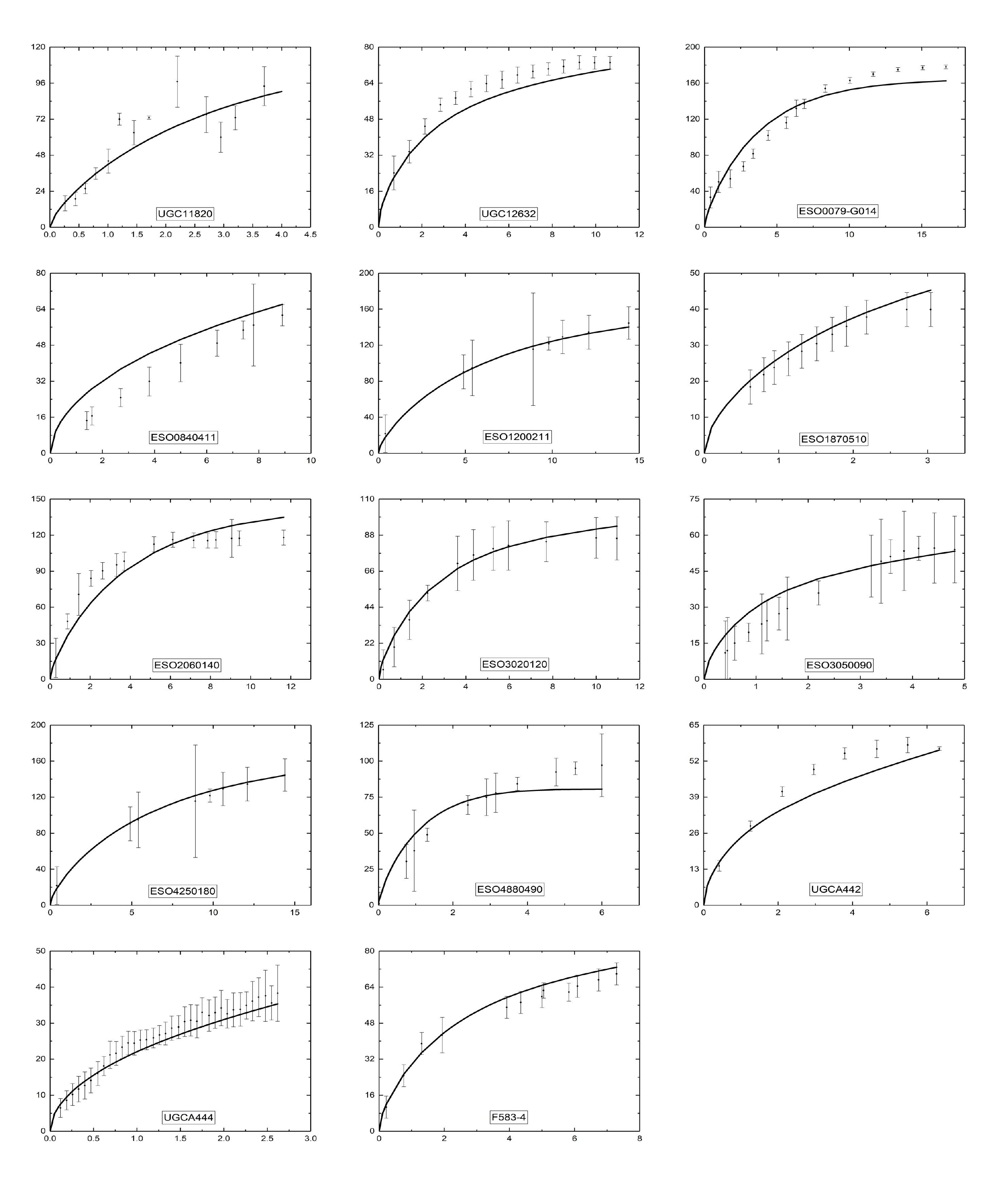}
	\caption{Fitting for the galaxies: UGC 11820, UGC 12632, ESO 079-G014, ESO 0840411, ESO 1200211, ESO 1870510, ESO 2060140, ESO 3020120, ESO 3050090, ESO 4250180, ESO 4880490, UGC A442, UGC A444, and F583-4.}
	\label{fig:8}
\end{figure}

From the 
geometry of the spacetime we have extracted the local and global contributions to the velocity of a probe star 
and fixed the two free parameters considering the actual rotations' curves of a sample of $104$ galaxies. The results are shown in the Fig.s. \ref{fig:2}-\ref{fig:8}. 
The plots shows an extraordinary agreement between the theoretical model and observational data, although for the larger galaxies the value of the cosmological constant is too small to make the model perfectly compatible with the data. Indeed, for example in UGC$128$, DDO$170$, NGC$1003$, UGC$1230$, NGC$3198$, UGC$6614$ the velocity at large distance is too large. 
On the other hand, for example the data relative to the galaxies F$563-1$, F$568-3$, NGC$2403$, NGC$2841$, NGC$3621$, NGC$7331$, UGC$11748$ are in perfect agreement with our model. 
%
%
However, our model is not completely realistic since we have modeled the surface brightness distribution and the intergalactic gas with simple exponentials profiles. Moreover, we did not take into account the thickness of the galactic disk and the uncertainty on the galactic inclination angle. These issues will be addressed in future works.

Nevertheless, we consider the outcome of our work a good reason to further reflect on the dark matter problem. Indeed, it is quite surprising that we can so well explain the galactic rotation curves in Einstein's conformal gravity without need to add higher derivative terms or exotic matter in the Einstein's equations. 
However, we would honestly emphasize that the missing matter (or missing gravity) seems need in many areas of cosmology and modern astrophysics that we here did not touch.

\begin{table}[htb]

	\centering
	
	\caption{Galactic parameters of our sample}
	
	\label{Tab01}
	
	\begin{ruledtabular}
		\begin{tabular}{ccccccccccc}

			\toprule
			
			&&	Distance
			&	$L_B$ &	$R_0$(kpc)  &	$M_{HI}$ &	$M_{disk}$ &&	 \multicolumn{3}{c}{Data sources}  \\
			
			\cmidrule(r){9-11}
			
			Galaxy&Type &(Mpc)&($10^{10}L_{\bigodot}$) &(kpc)&$(10^{10}M_{\bigodot})$ &$(10^{10}M_{\bigodot})$ & $(M/L)_{stars}$ &    $L_B$      &  $R_0$   &   HI  \\

			\hline
			
			M 33 &HSB& 0.9 & 0.85 & 2.5 & 0.11 & 1.13 &1.33& \cite{mannheim2012fitting} &\cite{kent1987surface} &	\cite{mannheim2012fitting}  \\
			NGC 55 &LSB& 1.9 & 0.588 & 1.9 & 0.13 & 0.3 & 0.5&	\cite{puche1991hi} & \cite{puche1991hi} & \cite{puche1991hi}\\
			DDO 64& LSB& 6.8 & 0.015 & 1.3 & 0.02& 0.04&2.87& \cite{smailagic2003uv} & \cite{smailagic2003uv} & \cite{smailagic2004lorentz}\\
			UGC 128 &LSB& 64.6 & 0.597 & 6.9 & 0.73 & 2.75 &4.6& \cite{mannheim2012fitting} & \cite{ansoldi2007non} & \cite{ansoldi2007non}\\
			DDO 154 &LSB& 4.2 & 0.007 & 0.8 & 0.03 & 0.003 &0.45& \cite{walter2008things}& \cite{leroy2008star} & \cite{walter2008things}\\
			DDO 168 &LSB& 4.5 & 0.032 & 1.2 & 0.03 & 0.06 &2.03& \cite{mannheim2012fitting} &\cite{mannheim2012fitting} & \cite{mannheim2012fitting}\\
			DDO 170 &LSB& 16.6 & 0.023 & 1.9 & 0.09 & 0.05 &1.97& \cite{lake1990distribution} & \cite{lake1990distribution} &\cite{lake1990distribution}\\
			UGC 191 &LSB& 15.9 & 0.129 & 1.7 & 0.26 & 0.49 &3.81& \cite{van2000evolutionary} & \cite{van2000evolutionary} & \cite{van1997comparative}\\
			NGC 247 &LSB& 3.6 & 0.512 & 4.2 & 0.16 & 1.25 &2.43& \cite{carignan1985surface} & \cite{carignan1985surface}& \cite{carignan1990hi}\\
			NGC 300 &LSB& 2.0 & 0.271 & 2.1 & 0.08 & 0.65 &2.41& \cite{carignan1985surface} & \cite{carignan1985surface} & \cite{puche1990hi}\\
			UGC 477 &LSB& 35.8 & 0.871 & 3.5 & 1.02 & 1.00 &1.14& \cite{fisher1981neutral} & \cite{mannheim2012fitting}& \cite{fisher1981neutral}\\
			F563-1 & LSB & 46.8 & 0.14 & 2.9 & 0.29& 1.35 & 9.65 & \cite{de1997dark}& \cite{de1997dark}& \cite{de1996h}\\
			F563-V2 & LSB & 57.8 & 0.266 & 2.0 & 0.2 & 0.6 & 2.26 & \cite{de1997dark}& \cite{de1997dark}& \cite{de1996h}\\
			F568-2 & LSB & 80.0 & 0.351 & 4.2 & 0.3 & 1.2 & 3.43 & \cite{de1997dark}& \cite{de1997dark}& \cite{de1996h}\\
			F571-8 & LSB & 50.3 & 0.191 & 5.4 & 0.16 & 4.48 & 23.49 & \cite{de1996h}&\cite{de1997dark}& \cite{de1996h}\\
			F579-V1 & LSB & 86.9 & 0.557 & 5.2 & 0.21 & 3.33 & 5.98 & \cite{de1996h}& \cite{de1997dark}& \cite{de1996h}\\
			F583-1 & LSB & 32.4 & 0.064 & 1.6 & 0.18 & 0.15 & 2.32 & \cite{de1997dark} & \cite{de1997dark}& \cite{de1996h}\\
			F583-4 & LSB&50.8&0.096&2.8&0.06&0.31&3.25&\cite{de1997dark} & \cite{de1997dark}& \cite{de1996h}\\
			F730-V1 & LSB & 148.3 & 0.756 & 5.8 & & 5.95 & 7.87 & \cite{mannheim2012fitting}& \cite{mannheim2012fitting}& ...\\
			NGC 925 & LSB & 8.7 & 1.444 & 3.9 & 0.41 & 1.372 & 0.95 & \cite{walter2008things}&\cite{leroy2008star} & \cite{walter2008things}\\
			NGC 959 & LSB & 13.5 & 0.333 & 1.3 & 0.05 & 0.37 & 1.11 & \cite{fisher1981neutral}&\cite{esipov1991ubvr}&\cite{fisher1981neutral}\\
			NGC 1003 & LSB & 11.8 & 1.48 & 1.9 & 0.63 & 0.66 &0.45 & \cite{mannheim2012fitting}&\cite{broeils1992mass}& \cite{sanders1996published} \\
			UGC 1230 & LSB & 54.1 & 0.366 & 4.7 & 0.65 & 0.67 & 1.82 & \cite{de1997dark}&\cite{van1993star}&\cite{van1993star}\\
			UGC 1281 & LSB & 5.1 & 0.017 & 1.6 & 0.03 & 0.01 & 0.53 & \cite{van2000evolutionary}&\cite{de2002high}&\cite{swaters2002westerbork}\\
			UGC 1551 & LSB & 35.6 & 0.78& 4.2 & 0.44 & 0.16 & 0.2 &\cite{swaters2002westerbork}&\cite{de1996near}&\cite{swaters2002westerbork}\\
			NGC 1560 & LSB & 3.7 & 0.053 & 1.6 & 0.12 & 0.17 & 3.16 & \cite{mannheim2012fitting} & \cite{mannheim2012fitting}&\cite{mannheim2012fitting}\\
			UGC 2259 & LSB & 10.0 & 0.11 & 1.4 & 0.04 & 0.47 & 4.23 & \cite{carignan1988hi} & \cite{kent1987dark} & \cite{carignan1988hi}\\
			NGC 2403 & HSB & 4.3 & 1.647 & 2.7 & 0.46 & 2.37 & 1.44 & \cite{walter2008things}&\cite{wevers1986palomar}& \cite{walter2008things}\\
			IC 2574 & LSB & 4.5 & 0.345 & 4.2 & 0.19 & 0.098& 0.28 & \cite{walter2008things} &\cite{pasquali2008large}&\cite{walter2008things}\\
			NGC 2683 & HSB & 10.2 & 1.882 & 2.4 & 0.15&6.03& 3.20&\cite{casertano1991declining}&\cite{kent1985ccd}&\cite{sanders1996published}\\
			NGC 2841 & HSB & 14.1 & 4.742 & 3.5 & 0.86 & 19.552 & 4.12 & \cite{walter2008things}&\cite{mannheim2012fitting}&\cite{walter2008things}\\
			NGC 2903 & HSB & 9.4 & 4.088 & 3.0& 0.49 & 7.155 & 1.75&\cite{walter2008things}&\cite{wevers1986palomar}&\cite{walter2008things}\\
			NGC 2976 & LSB & 3.6& 0.201& 1.2& 0.01& 0.322&1.6& \cite{walter2008things}&\cite{simon2003high}&\cite{walter2008things}\\
			NGC 3031 & HSB & 3.7&3.187&2.6&0.38&8.662&2.72&\cite{walter2008things}&\cite{murphy2008connecting}&\cite{walter2008things}\\
			NGC 3109& LSB&1.5&0.064&1.3&0.06&0.02&0.35&\cite{carignan1985light}&\cite{carignan1985light}&\cite{jobin1990dark}\\
			NGC 3198&HSB & 14.1& 3.241&4.0&1.06&3.644&1.12&\cite{walter2008things}&\cite{wevers1986palomar}&\cite{walter2008things}\\
			NGC 3521&HSB&12.2& 4.769&3.3& 1.03&9.245&1.94&\cite{walter2008things}&\cite{leroy2008star}&\cite{walter2008things}\\
			NGC 3621& HSB& 7.4& 2.048&2.9&0.89&2.891&1.41&\cite{walter2008things}&\cite{de2008high}&\cite{walter2008things}\\
			NGC 3627&HSB& 10.2&3.70&3.1&0.10&6.622&1.79&\cite{walter2008things}&\cite{leroy2008star}&\cite{walter2008things}\\
			NGC 3716& HSB&17.4&3.34&3.2&0.60&3.82&1.15&\cite{sanders1998rotation}&\cite{tully1996ursa}&\cite{sanders1998rotation}\\
			NGC 3769&HSB&15.5&0.684&1.5&0.41&1.36&1.99&\cite{sanders1998rotation}&\cite{tully1996ursa}&\cite{sanders1998rotation}\\
			NGC 3877&HSB&15.5&1.948&2.4&0.11&3.44&1.76&\cite{sanders1998rotation}&\cite{tully1996ursa}&\cite{sanders1998rotation}\\
			NGC 3893&HSB&18.1&2.928&2.4&0.59&5&1.71&\cite{sanders1998rotation}&\cite{tully1996ursa}&\cite{sanders1998rotation}\\
			NGC 3917&LSB&16.9&1.334&2.8&0.17&2.23&1.67&\cite{sanders1998rotation}&\cite{tully1996ursa}&\cite{sanders1998rotation}\\
			NGC 3949&HSB&18.4&2.327&1.7&0.35&2.37&1.02&\cite{sanders1998rotation}&\cite{tully1996ursa}&\cite{sanders1998rotation}\\
			NGC 3953&HSB&18.7&4.236&3.9&0.31&9.79&2.31&\cite{sanders1998rotation}&\cite{tully1996ursa}&\cite{sanders1998rotation}\\
			NGC 3972&HSB&18.6&0.978&2&0.13&1.49&1.53&\cite{sanders1998rotation}&\cite{tully1996ursa}&\cite{sanders1998rotation}\\
			NGC 3992&HSB&25.6&8.456&5.7&1.94&13.94&1.65&\cite{sanders1998rotation}&\cite{tully1996ursa}&\cite{sanders1998rotation}\\
			NGC 4010&LSB&18.4&0.883&3.4&0.29&2.03&2.30&\cite{sanders1998rotation}&\cite{tully1996ursa}&\cite{sanders1998rotation}\\
			NGC 4013&HSB&18.6&2.088&2.1&0.32&5.58&2.67&\cite{sanders1998rotation}&\cite{tully1996ursa}&\cite{sanders1998rotation}\\
			NGC 4051&HSB&14.6&2.281&2.3&0.18&3.17&1.39&\cite{sanders1998rotation}&\cite{tully1997ursa}&\cite{sanders1998rotation}\\
			NGC 4085&HSB&19.0&1.212&1.6&0.15&1.34&1.11&\cite{sanders1998rotation}&\cite{tully1996ursa}&\cite{sanders1998rotation}\\

			\bottomrule
		\end{tabular}
	\end{ruledtabular}
	
\end{table}
\newpage

\begin{table}[htb]

	\centering
	
	\caption{Galactic parameters of our sample}
	
	\label{Tab02}
	\begin{ruledtabular}
		
		\begin{tabular}{ccccccccccc}
			
			\toprule
			
			&&	Distance
			&	$L_B$ &	$R_0$(kpc)  &	$M_{HI}$ &	$M_{disk}$ &&	 \multicolumn{3}{c}{Data sources}  \\
			
			\cmidrule(r){9-11}
			
			Galaxy&Type &(Mpc)&($10^{10}L_{\bigodot}$) &(kpc)&$(10^{10}M_{\bigodot})$ &$(10^{10}M_{\bigodot})$ & $(M/L)_{stars}$ &    $L_B$      &  $R_0$   &   HI  \\
			
			\hline
			NGC 4088&HSB&15.8&2.957&2.8&0.64&4.67&1.58&\cite{sanders1998rotation}&\cite{tully1996ursa}&\cite{sanders1998rotation}\\
						NGC 4100&HSB&21.4&3.388&2.9&0.44&5.74&1.69&\cite{sanders1998rotation}&\cite{tully1996ursa}&\cite{sanders1998rotation}\\

			UGC 4115& LSB&5.5&0.004&0.3&&0.01&0.97&\cite{de2001high}&\cite{barazza2001structure}&...\\
			NGC 4138&LSB&15.6&0.827&1.2&0.11&2.97&3.59&\cite{sanders1998rotation}&\cite{tully1996ursa}&\cite{sanders1998rotation}\\
			NGC 4157&HSB&18.7&2.901&2.6&0.88&5.83&2.01&\cite{sanders1998rotation}&\cite{tully1996ursa}&\cite{sanders1998rotation}\\
			NGC 4183&HSB&16.7&1.042&2.9&0.30&1.43&1.38&\cite{sanders1998rotation}&\cite{tully1996ursa}&\cite{sanders1998rotation}\\
			NGC 4217&HSB&19.6&3.031&3.1&0.30&5.53&1.83&\cite{sanders1998rotation}&\cite{tully1996ursa}&\cite{sanders1998rotation}\\
			UGC 4325&LSB&11.9&0.373&1.9&0.10&0.40&1.08&\cite{swaters2002westerbork}&\cite{de2002high}&\cite{swaters2002westerbork}\\
			NGC 4389&HSB&15.5&0.61&1.2&0.04&0.42&0.68&\cite{sanders1998rotation}&\cite{tully1996ursa}&\cite{sanders1998rotation}\\
			NGC 4395& LSB&4.1&0.374&2.7&0.13&0.83&2.21&\cite{swaters2002westerbork}&\cite{de2002high}&\cite{swaters2002westerbork}\\
			NGC 4826&HSB&10.2&3.7&3.1&0.10&6.622&1.79&\cite{walter2008things}&\cite{leroy2008star}&\cite{walter2008things}\\
			
			UGC 5005&LSB&51.4&0.2&4.6&0.28&1.02&5.11&\cite{de1997dark}&\cite{van1993star}&\cite{van1993star}\\
			NGC 5585&HSB&9.0&0.333&2.0&0.28&0.36&1.09&\cite{cote1991dark}&\cite{cote1991dark}&\cite{cote1991dark}\\
			UGC 5716& Sm & 21.3& 0.0588 & 1.14&0.1094&0.0588&1.0&\cite{lelli2016sparc}&\cite{lelli2016sparc}&\cite{lelli2016sparc}\\
			UGC 5750& LSB&56.1&0.472&3.3&0.10&0.10&0.21&\cite{de1997dark}&\cite{van1993star}&\cite{van1993star}\\
			UGC 5999& LSB&44.9&0.17&4.4&0.18&3.36&19.81&\cite{de1997dark}&\cite{van1993star}&\cite{van1993star}\\
			UGC 6399& LSB& 18.7&0.291&2.4&0.07&0.59&2.04&\cite{sanders1998rotation}&\cite{tully1996ursa}&\cite{sanders1998rotation}\\
			UGC 6446 &LSB&15.9&0.263&1.9&0.24&0.36&1.36&\cite{sanders1998rotation}&\cite{tully1997ursa}&\cite{sanders1998rotation}\\
			NGC 6503&HSB&5.5&0.417&1.6&0.14&1.53&3.66&\cite{mannheim2012fitting}&\cite{wevers1986palomar}&\cite{mannheim2012fitting}\\
			UGC 6614& LSB&86.2&2.109&8.2&2.07&9.70&4.60&\cite{de2001high}&\cite{van1993star}&\cite{van1993star}\\
			UGC 6667&LSB&19.8&0.422&3.1&0.10&0.71&1.67&\cite{sanders1998rotation}&\cite{tully1996ursa}&\cite{sanders1998rotation}\\
			UGC 6818&LSB&21.7&0.352&2.1&0.16&0.11&0.33&\cite{sanders1998rotation}&\cite{tully1996ursa}&\cite{sanders1998rotation}\\
			UGC 6917&LSB&18.9&0.563&2.9&0.22&1.24&2.20&\cite{sanders1998rotation}&\cite{tully1996ursa}&\cite{sanders1998rotation}\\
			UGC 6923& LSB&18.0&0.297&1.5&0.08&0.35&1.18&\cite{sanders1998rotation}&\cite{tully1997ursa}&\cite{sanders1998rotation}\\
			UGC 6930& LSB&17.0&0.601&2.2&0.29&1.02&1.69&\cite{sanders1998rotation}&\cite{tully1996ursa}&\cite{sanders1998rotation}\\
			NGC 6946&HSB&6.9&3.732&2.9&0.57&6.272&1.68&\cite{walter2008things}& \cite{leroy2008star} & \cite{walter2008things}\\
			UGC 6973& HSB&25.3&1.647&2.2&0.35&3.99&2.42&\cite{sanders1998rotation}&\cite{tully1997ursa}&\cite{sanders1998rotation}\\
			UGC 6983& LSB&20.2&0.577&2.9&0.37&1.28&2.22&\cite{sanders1998rotation}&\cite{tully1996ursa}&\cite{sanders1998rotation}\\
			UGC 7089& LSB& 13.9&0.352&2.3&0.07&0.35&0.98&\cite{sanders1998rotation}&\cite{tully1996ursa}&\cite{sanders1998rotation}\\
			NGC 7137&LSB& 25.0& 0.959&1.7&0.10&0.27&0.28&\cite{bieging1977radio}&\cite{mannheim2012fitting}&\cite{bieging1977radio}\\
			NGC 7331& HSB& 14.2&6.773&3.2&0.85&12.086&1.78&\cite{walter2008things}& \cite{leroy2008star} & \cite{walter2008things}\\
			NGC 7793& HSB& 5.2&0.91&1.7&0.16&0.793&0.87&\cite{walter2008things}& \cite{leroy2008star} & \cite{walter2008things}\\
			UGC 11454 & LSB& 93.9&0.456&3.4&&3.15&6.90&\cite{de2001high}&\cite{mannheim2012fitting}&...\\
			UGC 11557& LSB&23.7&1.806&3.0&0.25&0.37&0.20&\cite{de2001high}&\cite{swaters2002westerbork}&\cite{swaters2002westerbork}\\
			UGC 11583& LSB& 7.1& 0.012&0.7&& 0.01&0.96&\cite{de2001high}&\cite{mannheim2012making}&...\\
			UGC 11616& LSB& 74.9&2.159&3.1&&2.43&1.13&\cite{de2001high}&\cite{mannheim2012making}&...\\
			UGC 11648& LSB& 49.0&4.073&4.0&&2.57&0.63&\cite{de2001high}&\cite{mannheim2012making}&...\\
			UGC 11748& LSB& 75.3&23.930& 2.6&& 9.67&0.40& \cite{de2001high}&\cite{mannheim2012making}&...\\
			UGC 11819& LSB& 61.5&2.155&4.7&&4.83&2.24&\cite{de2001high}&\cite{mannheim2012making}&...\\
			UGC 11820& LSB& 17.1&0.169&3.6&0.40&1.68&9.95&\cite{van1997comparative}&\cite{mannheim2012fitting}&\cite{van1997comparative}\\
			UGC 12632& Sm& 9.77&0.1301& 2.42&0.1744& 0.1301& 1.0&\cite{lelli2016sparc}&\cite{lelli2016sparc}&\cite{lelli2016sparc}\\
			ESO 079-G014 & Sbc& 28.7& 5.1733&5.08&0.314&5.1733& 1.0&\cite{lelli2016sparc}&\cite{lelli2016sparc}&\cite{lelli2016sparc}\\
			ESO 0840411& LSB & 82.4& 0.287& 3.5& & 0.06& 0.21& \cite{de2001high}&\cite{mannheim2012fitting}&...\\
			ESO 1200211& LSB& 15.2& 0.028& 2.0& & 0.01& 0.20& \cite{de2001high}&\cite{mannheim2012fitting}&...\\
			ESO 1870510& LSB& 16.8& 0.054& 2.1& & 0.09& 1.62& \cite{de2001high}&\cite{bell2000star}&...\\
			ESO 2060140& LSB& 59.6& 0.735& 5.1& & 3.51& 4.78& \cite{de2001high}&\cite{beijersbergen1999surface}&...\\
			ESO 3020120& LSB & 70.9& 0.717& 3.4& &0.77&1.07& \cite{de2001high}&\cite{mannheim2012fitting}&...\\
			ESO 3050090& LSB& 13.2& 0.186& 1.3& &0.06& 0.32& \cite{de2001high}&\cite{mannheim2012fitting}&... \\
			ESO 4250180& LSB& 88.3& 2.6& 7.3& & 4.79& 1.84& \cite{de2001high}&\cite{beijersbergen1999surface}&...\\
			ESO 4880490& LSB& 28.7& 0.139& 1.6& & 0.43& 3.07& \cite{de2001high}&\cite{mannheim2012fitting}&...\\
			UGC A442& Sm& 4.35& 0.014& 1.18& 0.0263& 0.014& 1.0& \cite{lelli2016sparc}&\cite{lelli2016sparc}&\cite{lelli2016sparc}\\
			UGC A444 & Im& 0.98& 0.0012& 0.083& 0.0067& 0.0012& 1.0& \cite{lelli2016sparc}&\cite{lelli2016sparc}&\cite{lelli2016sparc}\\

			\bottomrule
			
		\end{tabular}
	\end{ruledtabular}
\end{table}

\bibliographystyle{unsrt}
\bibliography{hept}
\end{document}